\documentstyle[twocolumn,eqsecnum,aps,epsfig]{revtex}

\begin{document}
\title{Proton and deuteron distributions as signatures
for collective particle dynamics and event shape geometries 
at the RHIC}
\author{B. Monreal$^{(a),\dag}$, W.J. Llope$^{(b)}$,
  R. Mattiello$^{(c)}$, S.Y. Panitkin$^{(d)}$, H. Sorge$^{(e)}$, N. Xu$^{(a)}$
}  
\address{\em (a) Nuclear Science Division, LBNL, Berkeley, CA 94720,
USA} \address{\em (b) T.W.~Bonner Nuclear Laboratory,
Rice University, TX 77005, USA} \address{\em (c) Niels
Bohr Institute, Blegdamsvej 17, University of Copenhagen, DK-2100
Copenhagen, Denmark} \address{\em (d) Department of 
Physics, Kent State University, OH 44242, USA} \address{\em (e)
Department of Physics, SUNY at Stony Brook, NY 11794, USA} 
\date{\today}
\maketitle
\begin{abstract}
We present predictions for the formation of 
(anti)nuclear bound states in nucleus-nucleus reactions
at RHIC energies.
The phase space coalescence method is used
in combination with RQMD-v2.4 transport calculations to demonstrate 
the relevance of particle production as well as
the longitudinal and transverse flow components.
The formation of deuterons follows an
approximate scaling law proportional
to the relative freeze-out densities of 
nucleons and produced secondaries.  
For antideuterons, an additional suppression
appears that is proportional to the number of nucleons,
pointing toward multiple rescattering and absorption
prior to freeze-out.
\end{abstract}
\pacs{PACS numbers: 25.75.-q, 25.75.Ld, 25.75.Dw, 24.10.Lx}
%
%
\narrowtext
Nuclear clusters have been a useful tool to establish
collective effects throughout the history of heavy ion reactions.
Their production rates
have provided evidence for low temperature
phase transitions \cite{ALADIN}, their spectral distribution shows
particular sensitivities to collective flow \cite{Lisa,E802,E877},
transverse expansion \cite{UHeinz,PRL,Polleri,Nagle} and potential
forces \cite{PRL,Danielewiczs}. 
In case of strong enough ``cooling'' of the emitting source and
collective motion even the study of bound states with a considerable
fraction  of antimatter \cite{Bleicher}, strangeness or even charm
\cite{Schaff} becomes possible. 
Light antimatter clusters up to $A=3$ have already been identified
\cite{QM96}, while the search for states with strange constituents
is ongoing.
Deuterons and antideuterons are the 
simplest composite objects and are useful in
establishing expansion and
correlations in the emitting source.
Volume-expansion due to secondary interactions
tends to diminish the cluster yields as particle
production rises both with the beam energy and the system size \cite{Sorge1}.
Counterbalancing effects can be expected
from collective flow components that  increase
 cluster multiplicities and
reduce effective source radii as compared to
the actual size of the system \cite{PRL}.
It should be emphasized that the predictions of different transport models
for collisions at RHIC energies already offer large differences
in the most basic observables. For example, the total
number of pions predicted at midrapidity varies by factors
$\approx$2 in comparisons with parton cascade and RQMD-type
calculations~\cite{RHIC98}.
Constraints from nuclear bound state analyses ({\it i.e.} fragment
production) should complement those
from single inclusive hadron spectra
and pion/proton interferometry in order to
distinguish the different model scenarios.
Moreover, collective motion,
temperatures and position densities are related
to entropy production and pressures, major
assets in the search for QCD-phase transitions.
We therefore suggest cluster analyses will remain a relevant tool in
the upcoming experiments at RHIC. 
In this paper we present predictions  for (anti)proton and
(anti)deuteron production
based on the transport approach
RQMD v2.4~\cite{Sorge95} in combination with a
phase space coalescence framework~\cite{PRL,Nagle,Sorge1}.
The RQMD-transport model is a semiclassical 
microscopic approach which combines classical propagation
with stochastic interactions. Color strings and hadronic
resonances can be excited in elementary collisions.
Their fragmentation and decay lead to particle production.
Overlapping strings may form ropes, chromoelectric
flux tubes with charges in higher dimensional
representations of color SU(3). RQMD is a full
transport theoretical approach to reactions between nuclei
and elementary hadrons. Since the model does not include
light cluster production it is  supplemented with a coalescence
``afterburner.''  
The coalescence is performed by projecting the 
2-body phase space density given by the microscopic transport
onto bound state
wave functions in Wigner-space. This method has been
successful in the description of deuteron formation
and proton-deuteron correlations at lower beam energies 
\cite{PRL,Nagle,Panitkin1}.
The basic observables that demonstrate phase space sensitivities 
are the rapidity and transverse momentum distributions.
Figure 1 (upper panels) shows predictions 
for (anti)protons  and (anti)deuterons 
in central Au+Au collisions at full RHIC energy\footnote{$100
A\cdot GeV Au + 100 A\cdot GeV Au$, impact parameters $0 \leq b \leq 3$, with 
only participant nucleons considered.}. 
Fig. 1 (lower panels) presents the rapidity dependence of average  
transverse momenta. The values exhibit approximately
a factor of two higher transverse momentum in the
composite objects. 
The average transverse momenta
of antiprotons(deuterons) are slightly larger than
those of their matter counterparts.
The predictions for particles and antiparticles shown in Fig. 1
are summarised in Tables 1 and 2. 
The presence of flow components
can be demonstrated by comparing
with calculations
where the freeze-out correlations of positions and momenta have
been removed by hand via randomization (crosses on Fig. 1).
As a result of the randomization process the
total number of clusters at midrapidity is
suppressed by orders of magnitude.
The average transverse momenta at midrapidity in the composite
spectra drop by approximately 30\%, consistent with simple momentum
coalescence results. 
The collective component in the transverse momentum of
deuterons and antideuterons is 
correlated with the number of
scatterings: Collective expansion is highest
in the midrapidity region
while it becomes less pronounced
close to projectile and target rapidities.
A more detailed investigation
of transverse flow issues
can be found in Ref.~\cite{Monreal1}.
In the following, ratios of particle
yields are examined as measures of collectivity
and (relative) freeze-out densities in longitudinal
momentum.
The particle yields show relatively
small variations in rapidity space
and the system shows strong evidence of a ``collective''
evolution towards freeze-out.
In this case, the nuclear phase space densities 
can be approximated
by average position
densities and local momentum 
fluctuations, usually addressed as ``temperatures''.
For a purely thermal source the $N_d/N_p$-ratio 
reflects the nuclear density
$N_d/N_p\propto N_p/V\sim\langle\rho_p\rangle$ (where $N_d$ is number of
deuterons, $N_p$ is  number of protons, $\rho_p$ is proton freeze-out
density, V- freeze-out volume)  
and should not depend on the particle velocities.
The transport calculations strongly deviate from  
such a scenario (see Fig. 2, a and c) showing
$N_d/N_p$- and $N_d/N_p^2$- ratios that strongly vary 
as function of rapidity.  
This behaviour can be traced back to strong longitudinal flow
components that lead to a partial separation of sources in beam
direction. 
The effective ``volume'', $V_p$ is proportional to the ratio, $\propto
p^2/d$, but seems to be enlarged proportional to the number of
secondaries $N_{sec}$ (here $N_{sec}$ is a number of produced mesons:
$N_{sec} = N(\pi^{\pm,0})+N(K^{\pm,0}$)) 
thus supporting the role of particle production
and rescattering.
This can be demonstrated by regarding the ``scaled''
ratio $(N_d/N_p^2)*N_{sec}$ for which
the differences basically vanish.
In semicentral reactions less expansion  
is found in the $N_d/N_p^2$ ratio while
the scaled ratio turns out to be rather similar.
The comparison of reactions with different
projectile or target size has been found  useful
to assess the relative strength of 
hadronic expansion at SPS-energies \cite{Sorge1}. 
Such size dependences are not seen in our centrality dependence
analysis although the effective nuclear volume seems to differ by factors 
3-5. 
Note  that the naive density-interpretation of the $N_d/N_p$ ratio
can be somewhat flawed by transverse
and directed flow correlations which change with rapidity.
Antideuterons and antiprotons reveal
differing emissions patterns that are dominated by the
source center. The $N_{\bar{d}}/N_{\bar{p}}$ ratio
(see Fig. 2, b) is rather flat and 
could lead to misleading conclusions
since flow correlations $appear$ to be close to none. 
The presence of flow expresses itself in  
the scaled and unscaled ratios $N_{\bar{d}}/N_{\bar{p}}^2$ and
$(N_{\bar{d}}/N_{\bar{p}}^2) * N_{sec}$.
They are only consistent with a volume-type scaling
close to midrapidity but otherwise deviate
considerably.
This deviation can be explained
by nuclear absorption which is larger due to the presence
of higher nuclear densities and lower numbers of secondaries, 
particularly in the
domain $|y|>3$.
As a consequence, the differences become most prominent
in semi-central reactions with less scattering of baryons
towards midrapidity.
One further suggestion to address antimatter-absorption
has been suppression at low transverse
momenta as well as correlations in longitudinal
and transverse flow (event plane asymmetries)~\cite{Bleicher}.
In agreement with such a scenario, the average transverse
momenta for antinucleons are slightly larger than those for nucleons. 
It has been suggested~\cite{Llope} that
the ``homogeneity volume'' deduced from cluster production
should be compared to
radius parameters extracted by HBT-type two-particle correlation analysis. 
We would like to point out that such an analysis
suffers from major uncertainties
which are related to the widths
of nuclear clusters in position and momentum space.
Unlike HBT-type analysis, where the correlation strength
can be scanned as a function of the relative momentum
of particle pairs ($q_l,q_s,q_o$),
the projection on boundstate wave functions involves
both integrations over relative momenta and positions.
Hence, cluster analyses are more suitable to study
the $average$ phase space volume. 
Details of the emitting source in position space  
cannot uniquely be addressed
unless the size and shape of the (local) momentum fluctuations are
known.  
A solution to this caveat
could be  the study of various cluster types
such as
deuterons and $^4$He with 
wave functions (correlations) very different in position and momentum space.
Heavier clusters, in addition, provide less
sensitivity to local momentum fluctuations than the
loosely bound deuteron state.
As collective flow becomes strong enough,
the characteristic scaling of the relative
yields can give access to the flow and density geometry
(``event shape'') \cite{Polleri1}.
Further insight can be expected from proton-proton and proton-deuteron
correlation analysis (see for example \cite{Panitkin1}) which should
confirm the shape of deuteron phase space densities.
In summary, using the transport model RQMD(v2.4) and a coalescence
afterburner that projects RQMD's two body phase space densities onto
fragment wave functions in Wigner space, we study the production of
nucleons and 
deuterons (and their antiparticles) for the central 100A$\cdot$GeV Au +
100A$\cdot$GeV Au collisions. Rapidity distributions
and average transverse momentum  exhibit
strong longitudinal and transverse flow components.
As a consequence,
composite ratios are closely related to the 
position space distributions of the nucleons and
produced hadrons close to freeze-out.
Deuteron formation is consistent with a scaling
relationship proportional to the relative densities of nucleons
and secondaries at similar rapidities.
The spectra of antideuterons
are strongly modified during the course of the reaction.
Contributions from antimatter absorption
lead to deviations of antideuteron production from the naive 
volume scaling and slightly higher transverse momenta
as compared to their matter counterparts. 
Yet to be addressed are observables such as
elliptic~\cite{Olitreau,Sorge2} or directed~\cite{PRL,NA49} flow
patterns which are more sensitive to details  
in the early and late event shape.
 We are grateful for many enlightening discussions with
 Drs. S. Johnson, D. Keane, S. Pratt, H.G. Ritter, S. Voloshin. 
This research used resources of the National
 Energy Research Scientific Computing Center.  This work has been
 supported by the U.S. Department of Energy under Contract
 No. DE-AC03-76SF00098 and W-7405-ENG-36, the Energy Research 
 Undergraduate Laboratory Fellowship and National Science
 Foundation and the Marie-Curie Research Training 
Grant No. FMBICT961721.

$\dag$ B. Monreal is at Lawrence Berkeley National Laboratory through
the Center for Science and Engineering Education.\\


\begin{table}
\begin{tabular}{c|c|c|c|c}
Rapidity & p & d & $\bar{p}$ & $\bar{d}$ \cr \hline
   0.25 &  15.6 &   0.042 &  7.4 &   0.0092 \\
   0.75 &  15.9 &   0.049 &  7.1 &   0.0098 \\
   1.25 &  16.2 &   0.057 &  7.0 &   0.0109 \\
   1.75 &  16.9 &   0.078 &  6.6 &   0.0093 \\
   2.25 &  18.9 &   0.098 &  5.9 &   0.0095 \\
   2.75 &  22.0 &   0.150 &  5.1 &   0.0068 \\
   3.25 &  22.9 &   0.195 &  4.5 &   0.0064 \\
   3.75 &  21.8 &   0.202 &  4.1 &   0.0057 \\
   4.25 &  17.5 &   0.208 &  2.9 &   0.0049 \\
   4.75 &  11.3 &   0.159 &  1.0 &   0.0012 \cr
\end{tabular}
\caption{Predicted production yield for protons, deuterons,
antiprotons and antideuterons as a function of rapidity.}    

\label{Tab:rapidities}
\end{table}

\begin{table}
\begin{tabular}{c|c|c|c|c}
Rapidity & p & d & $\bar{p}$ & $\bar{d}$ \cr \hline
   0.25 &   0.89 &   0.69 &    0.93 &   0.72 \\
   0.75 &   0.90 &   0.67 &    0.97 &   0.73 \\
   1.25 &   0.88 &   0.67 &    0.96 &   0.71 \\
   1.75 &   0.87 &   0.65 &    0.93 &   0.68 \\
   2.25 &   0.84 &   0.63 &    0.92 &   0.66 \\
   2.75 &   0.80 &   0.59 &    0.86 &   0.58 \\
   3.25 &   0.73 &   0.53 &    0.80 &   0.59 \\
   3.75 &   0.66 &   0.45 &    0.73 &   0.47 \\
   4.25 &   0.58 &   0.37 &    0.62 &   0.35 \\
   4.75 &   0.49 &   0.30 &    0.54 &   0.40 \cr
\end{tabular}
\caption{Predicted mean transverse momenta (in GeV/c) for protons, deuterons,
antiprotons and antideuterons as a function of rapidity.}    
\label{Tab:mean_pt}
\end{table}
\begin{center}
\newpage
\begin{figure}[h]
\centerline{\epsfig{figure=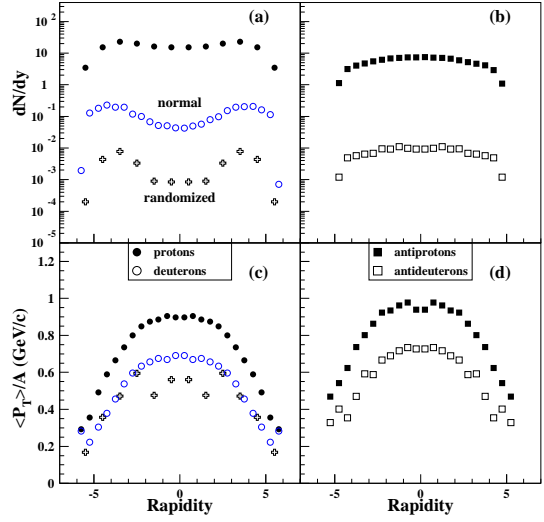,width=8.0cm}}
\caption{ Predicted rapidity distributions of {\bf (a)} protons
(filled circles), deuterons (open circles) and  
{\bf (b)} anti-protons (filled squares) and
anti-deuterons (open squares) for  
Au(200AGeV)Au reactions (b$<$3fm).
{\bf (c)} and {\bf (d)}  show the corresponding mean transverse
momenta per nucleon as function of rapidity for protons,  deuterons, 
antiprotons and antideuterons. 
Crosses indicate deuterons from calculations without position-momentum
correlations in the nuclear source.
}
\end{figure}
\begin{figure}[c]
\centerline{\epsfig{figure=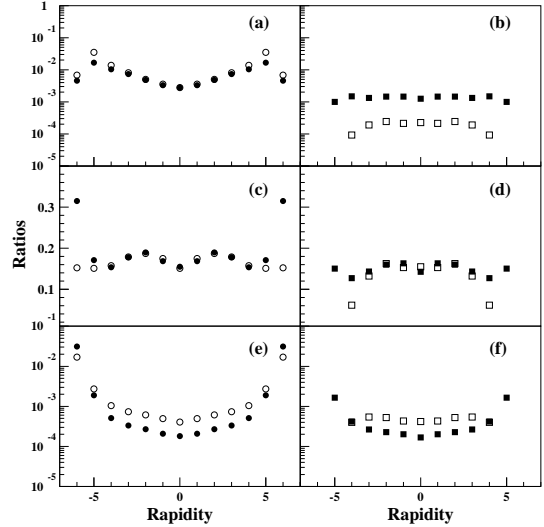,width=8.cm}}
\caption{Predicted ratio of particle yields as function of rapidity for  
 central (solid symbols) and semicentral (b$=$6fm) events (open symbols):
{\bf (a)-(b)} $N_d/N_p$ and $N_{\bar d}/N_{\bar p}$ ratios,
{\bf (c)-(d)} scaled ratios $(N_d/N_p^2)*N_{sec}$ and $(N_{\bar
d}/N_{\bar p}^2)*N_{sec}$   
and
{\bf (e)-(f)} $N_d/N_p^2$ and $N_{\bar d}/N_{\bar p}^2$ ratios.
}
\end{figure}

\end{center}  

\end{document}